\documentclass[11pt,a4paper]{article}
\usepackage[left=3.5 cm, right=3 cm, top=4 cm, bottom=4 cm]{geometry} 
\usepackage[pdfborder={0 0 0},plainpages=false,pdfpagelabels]{hyperref}
\usepackage[latin1]{inputenc}
\usepackage{amsmath}
\usepackage{amsfonts}
\usepackage{amssymb}
\usepackage{makeidx}
\usepackage{graphicx}
\usepackage{float}  
\usepackage{multicol} 
\begin{document}
\begin{center}
Accepted in International Conference on Neutrino Physics and Astrophysics (ICNPA 
2012), Amsterdam May 2012. \linebreak\linebreak \linebreak
\textbf{RESULTS OF LONG-BASELINE NEUTRINO EXPERIMENTS 
BASED ON THE CONSIDERATION OF FRAMEWORKS}
\linebreak \linebreak
Md. Farid Ahmed\\
York University, 4700 Keele Street, Toronto, Ontario, Canada-M3J 1P3. \\ E-mail: mdfarid@yorku.ca
\end{center}
\bigskip
\textbf{ABSTRACT:}
In this paper, we present an analysis of the outcomes of long-baseline neutrino experiments based on the consideration of frameworks. Our analysis suggests that the time difference between the time-of-flights corresponding to the speed of light (in vacuum) and that of the neutrino from a source to a detector is within the range of our predicted time difference assuming that the speed of neutrino follows a Galilean transformation. As the ghostly particles called neutrinos have peculiar properties which are still unresolved, so we propose measurements need to be performed at different times of the day to test any diurnal variations (if possible then any annual variations) due to the movement of the Earth. 
\bigskip\\
\textbf{Shortened version of the title:}\\
THE NEUTRINO EXPERIMENT AND FRAMEWORKS 
\section{Introduction} As we are in 2012, the 125th anniversary of the Michelson-Morley experiment, we note that the historical process leading to the establishment of the constancy of the speed of light - the fundamental postulate of Special Relativity (SR) - as one of the cornerstones of modern physics is not due to a single experiment, but rather to a series of experimental and theoretical developments since 1864 by the publication of James Clerk Maxwell's equations of electromagnetism [1, 2]. The experimental demonstration of the production and detection of electromagnetic waves by Heinrich Hertz in 1887[1, 2]; the Michelson-Morley experiment by Michelson and Morley in 1887 [3]; the Lorentz transformation by Lorentz in 1904 [4] and Poincare in 1905 [5] led to the establishment of the constancy of the speed of light.\\
\\In 1905, Albert Einstein introduced Special Relativity [6] - a theoretical framework that proved immediately successful in unifying Maxwell's electrodynamics with classical Mechanics. SR maintains that all inertial frames of reference are equivalent and the velocity of light is a universal constant in these inertial frames independent of the velocity of the source or the observer. \\
\\Most of today's fundamental physics theories are based on the constancy of the speed of light. This is one of the foundation blocks upon which modern physics is built. It is well known that the Standard Model of Particle Physics ensures that it must be consistent with SR. Despite the remarkable success of the theories based on the constancy of the speed of light, several modern theoretical approaches have begun to predict variations to the constant light-speed postulate. For example, string theory which seeks to unify today's Standard-Model with general relativity predicts a violation of the constancy of the speed of light [7 - 9].\\
\\ A few years back in 2007, the MINOS Collaboration [10] at Fermilab in USA reported for neutrinos moving faster than light. Also, recently in 2011, the Oscillation Project with Emulsion-Racking 
Apparatus (OPERA)-collaboration has reported [11] that "neutrinos travel faster than speed of light". There are a significant number of suggestions, arguments and counter arguments of the highly publicized preprint in arXiv of the OPERA's claim which are growing every day and produce the highest number of citations within the shortest period of time so far (100 citations within two weeks). However, the team (OPERA) has now found two problems - one in its timing gear 
and one in an optical fibre connection - that may have affected their tests (such as a report in BBC News on 23 February, 2012). Indeed, there are, also, other reasons suggested 
by different authors (such as Ref. [12]) which encourage severe scrutiny of the OPERA result. Therefore, as far as we know that the OPERA will perform more tests in near 
future to observe how they (timing gear and faulty connection) affect measured speeds of the neutrinos. Also other teams will and should perform the experiment to ensure 
that these possible errors account for the faster than speed of light results. However, if confirmed, this finding would overturn the most fundamental postulate of Einstein's 
Special Relativity that "nothing travels faster than 299,792,458 meters per second (the speed of light in vacuum)". The meaning of this outcome is simple to understand but 
the consequences would be far reaching. Let us look at the neutrino data of long-baseline neutrino experiments based on the consideration of frameworks and what they 
indicate.
\section{Frameworks}
In an ideal consideration for a frame transformation, we use rest and moving reference frames in rectilinear relative motion. We note that the moving inertial frame is replaced by the Earth frame in a terrestrial experimental investigation which is, indeed, not an inertial frame. The motion relative to Earth's centre of mass of a point on the equator of the Earth is about $5\times10^{2} m/s$. As well the Earth travels at a speed of approximately $3\times10^{4} m/s$ in its orbit around the Sun. Also the Sun is traveling together with its planets about the galactic centre with a speed $2.5\times10^{5} m/s$, and there are other motions at higher levels of the structure of the universe. Smoot et al [13] summarize the different velocities of our Solar system (the Earth) relative to the cosmic blackbody radiation, nearby galaxies and the Milky Way galaxy; also the motion of the Milky Way galaxy relative to the cosmic blackbody radiation. Therefore, the Earth experiences a significant motion relative to a rest frame, which is $\geq 600 km/sec$ [13]. An illustration of various movements of the Earth as well as its positions in different seasons (as for example on March 7 and September 7 in 2011) is shown in Fig. 1.\\
\\ We derived the time dependent components of the velocity of the laboratory along the direction of the light/neutrino propagation in our reports [14 - 16] assuming the Cosmic Microwave Background (CMB) is the rest frame of the universe. This derivation can help us to understand the shape of the change of velocity of the laboratory relative to the rest frame. As for example, following the propagation direction of neutrinos in the laboratory at CERN, the direction to the line CERN (source) - Gran Sasso (Detector) is roughly parallel to West to East [17], we derive the time dependent component of the velocity of the laboratory relative to the rest frame $[V(t)_{E-W-CMB}]$. 
Therefore:
\begin{figure}[t]
\includegraphics[width=1\textwidth]{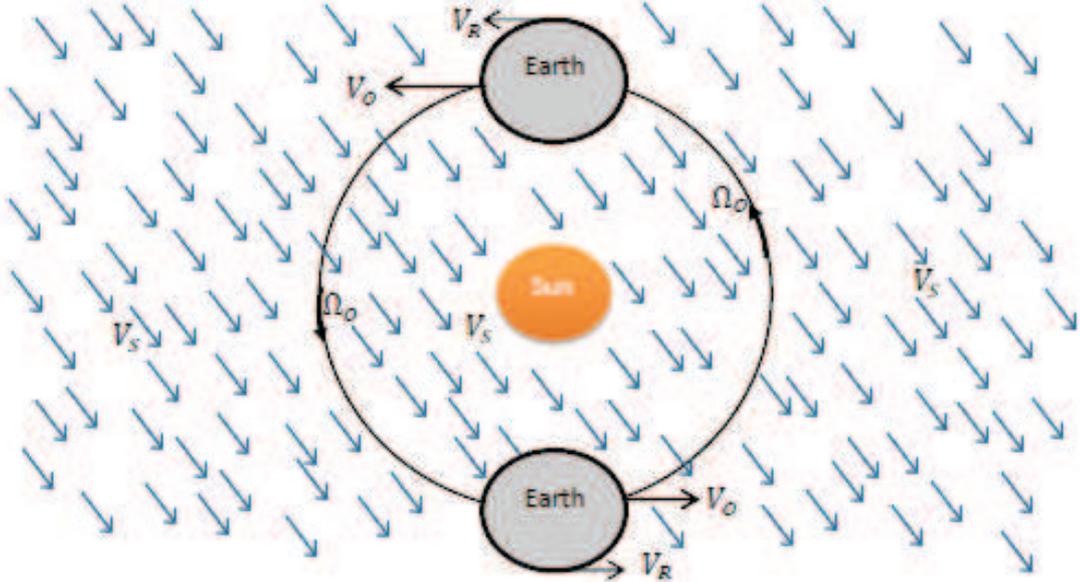} 
\caption{An illustration which represents various movements of the Earth as well as its positions in different seasons, as for example, on March 7 and September 7 in 2011.}
\end{figure}
\begin{equation}
\begin{split}
V(t)_{E-W-CMB}=&V_{O}[\sin(\Omega_{S}t)\sin(\Omega_{O}t)+\cos\varepsilon \cos(\Omega_{S}t) \cos(\Omega_{O}t)]   \\ 
&+V_{R}+V_{S}[\sin\alpha \cos\delta \sin(\Omega_{S}t)] 
\end{split}
\end{equation}
where $\chi=$co-latitude, $\alpha=$ Right Ascension, $\delta=$ Declination, $\varepsilon=$ the angle between the ecliptic and the Sun centered Celestial Equatorial plane, $\Omega_{S}=$ sidereal angular rotational frequency $(= 2\pi/(23h 56\min)\cong 4.18 \times 10^{-3}  deg. s^{-1})$, $V_{R}=$ the velocity due to the Earth's rotation about its axis depending on the geographical latitude $0\leq V_{R}\leq 4.5\times 10^{2} ms^{-1}$, $\Omega_{O}=$ the Earth is orbiting relative to the Sun with the angular frequency $(=2\pi/(1yr)\cong 1.14\times 10^{-5} deg.s^{-1})$, $V_{O}=$ the velocity due to the Earth's orbital motion relative to the Sun $(\approx 3\times 10^{4} m.s^{-1})$ and $V_{S}=$ the velocity of the solar system towards $(\alpha,\delta)$ relative to the rest frame. For the CMB as the rest frame we take $V_{S}=$ the velocity of the solar system towards $[(\alpha, \delta)=(168\deg, -7.22\deg)]$ relative to the CMB $(\approx 3.71 \times 10^{5} m.s^{-1})$ [12].\\
\\ Using equation (1), we derive the time dependent component of the velocity of the laboratory relative to the center of the solar system $[V(t)_{E-W}]$ as follows:
\begin{equation}
\begin{split}
V(t)_{E-W}=V_{O}[\sin(\Omega_{S}t)\sin(\Omega_{O}t)+\cos\varepsilon \cos(\Omega_{S}t) \cos(\Omega_{O}t)]+V_{R} 
\end{split}
\end{equation}
For our present discussion and analysis, we present $V(t)_{E-W-CMB}$ of equation (1) for a laboratory at CERN with respect to the CMB in Fig. 2. 
\begin{figure}[h]
\includegraphics[width=1\textwidth]{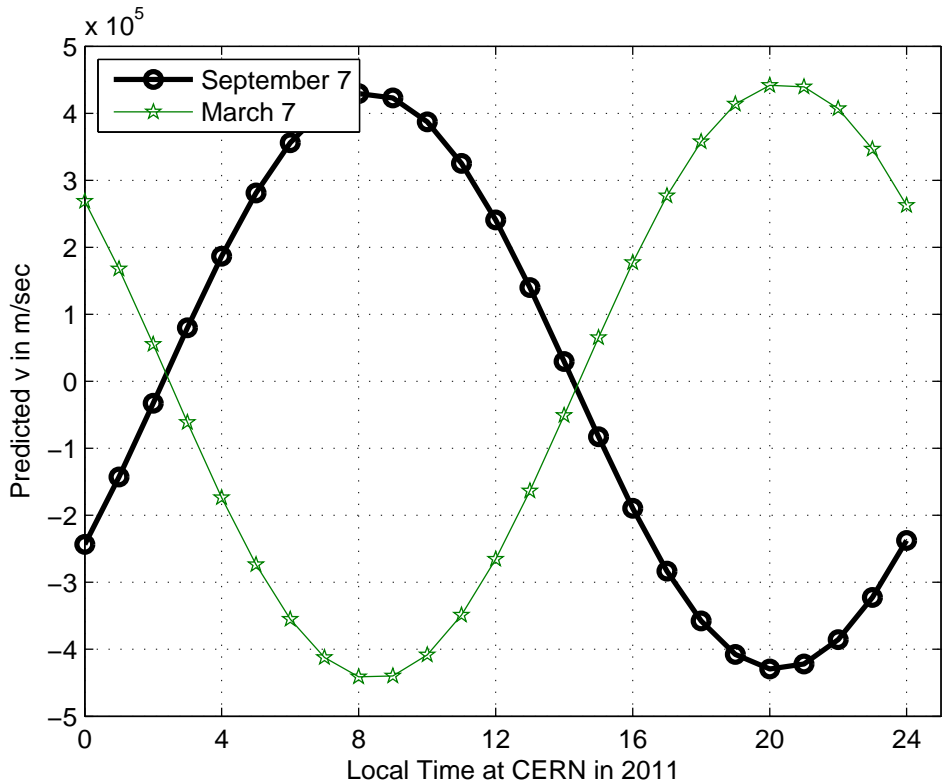} 
\caption{Presentation of the time dependent components of the velocities of a laboratory at the CERN $V(t)_{E-W-CMB}$ of equation (1)] with respect to the CMB along the East-West direction for the Earth's positions in different seasons as shown in Fig. 1.}
\end{figure}
\\
\\ Emission theories were proposed where electrodynamics was modified by supposing that the velocity of a light wave remained associated with the source rather than with a local or universal frame [18]. Following an emission theory, we assume that the speed of neutrino $(u)$ depends on the speed of the source $V(t)_{E-W-CMB}$ or $V(t)_{E-W}$ . 
\section{Long-baseline neutrino experiments}
Following Zuber [19], we know that there are three projects to perform long-baseline neutrino experiments to find out the speed of the neutrino. These projects are in Asia (sending a neutrino beam from a source at KEK to detectors at Super-Kamiokande in Japan with a baseline distance of about 250 km), North America (sending a neutrino beam from a source at Fermilab to a detector at the Soudan mine with a baseline distance about 730 km in USA) and Europe (sending a neutrino beam from a source at CERN to a detector at the Gran Sasso Laboratory with a baseline distance about 732 km). We present the outcomes of these long-baseline neutrino experiments, which were performed in different years, in Fig. 3. Here we choose the results of the neutrinos with average energy of $<40$ GeV as shown in Fig. 2 of Ref. [12] for our present analysis. The KAMIOKANDE-II Collaboration [21] data was obtained in Japan and reported in 1987. The Fermilab [I] data and the MINOS Collaboration [10] data were obtained at Fermilab in USA and reported in 1979 and 2007 respectively. The OPERA Collaboration [11] data was obtained in Europe and reported in 2011.  
\begin{figure}[h]
\includegraphics[width=1\textwidth]{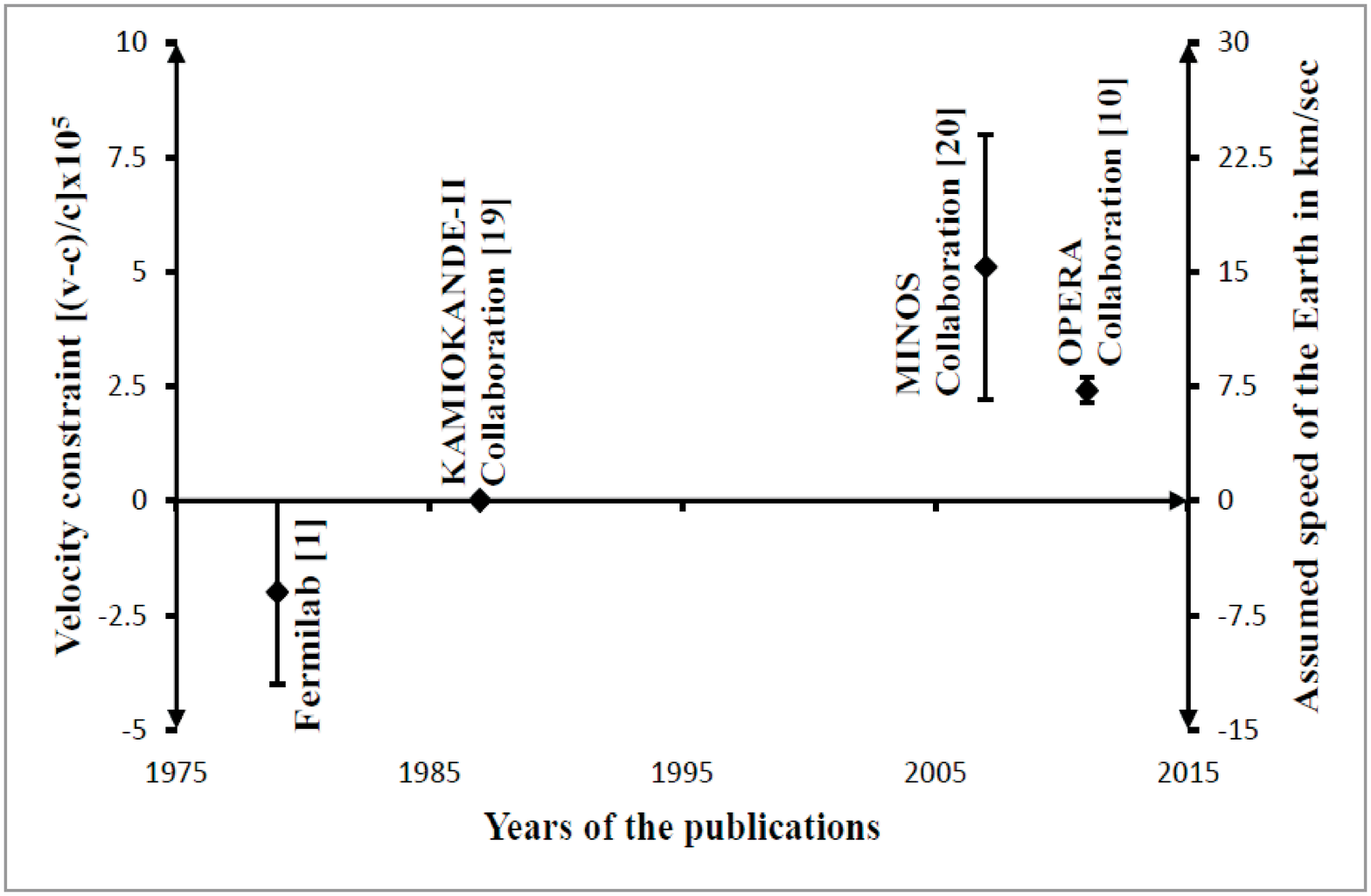} 
\caption{Outcomes of different long-baseline neutrino ($<40$ GeV) experiments where u is the velocity of the neutrino and c is the velocity of light in vacuum. [I] As in Fig. 2 of Ref. [12] and [20].}
\end{figure}
\\ 
\\We already noted in section 2 that for our present analysis and discussion, as for example, we look at the frameworks of long-baseline neutrino experiment of the OPERA Collaboration in detail. This experiment lies 1,400 meters underground in the Gran Sasso National Laboratory in Italy. It is designed to study a beam of neutrinos with average energy of $\sim17$ GeV coming from the CNGS neutrino beam at CERN (CERN Neutrino beam to Gran Sasso), Europe's premier high-energy physics laboratory located 732 kilometers away near Geneva, Switzerland. Let us note the parameters of this measurement following [11]: the speed of light in vacuum, $c\cong299,792,458 m/s$; the baseline used in the OPERA measurement, $L\cong 732,000 m$; the velocity of the neutrino, $u$; the time of flight corresponding to $c$, $L/c$; the time of flight corresponding to $u$, $L/u$. \\
\\
Therefore, the difference is $(\frac{L}{c} - \frac{L}{u})=(60\pm6.9(stat.)\pm7.4(sys.)) nanosecond$.\\
\\
Fig. 4 presents the predicted differences between time-of-flights corresponding to $c$ and $u$ from CERN to LNGS $(\dfrac{L}{c+V(t)_{E-W-CMB}} -\dfrac{L}{c} )$ assuming a Galilean transformation for the speed of neutrino. We assume that the speed of neutrino is $u=c + V(t)_{E-W-CMB}$.    
\begin{figure}[h]
\includegraphics[width=1\textwidth]{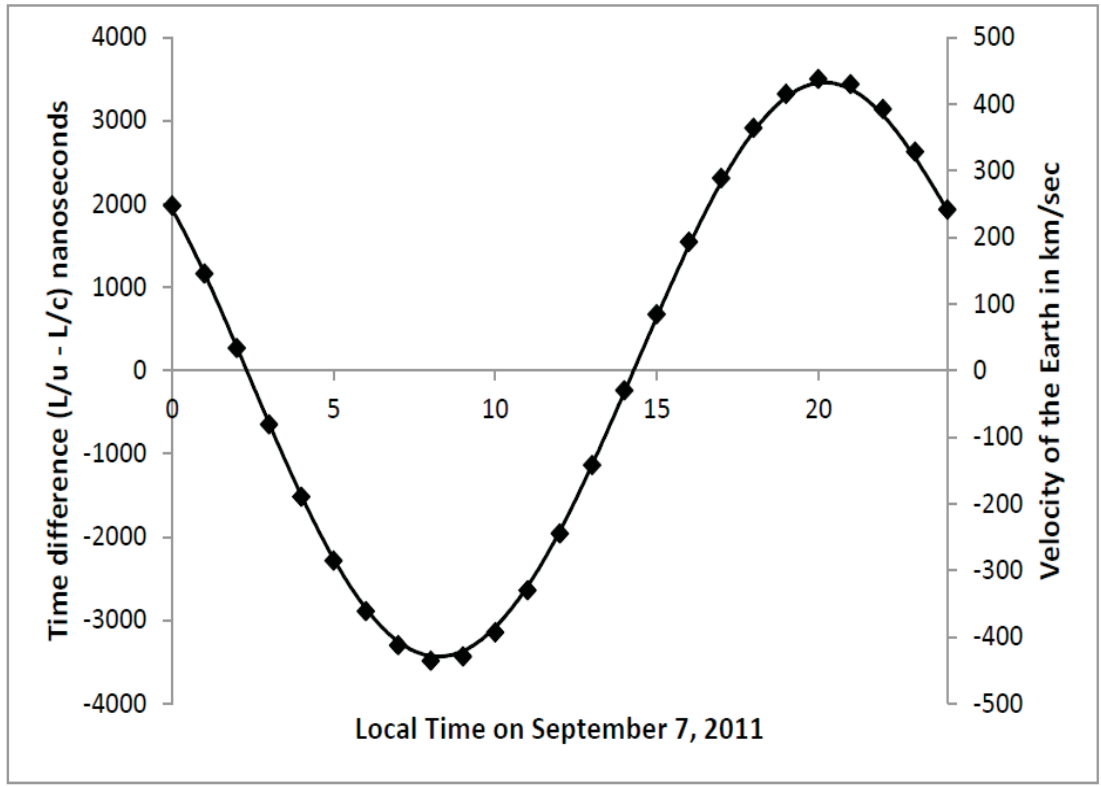} 
\caption{Presentation of a predicted time difference between the time-of-flights corresponding to $c$ and $u$ from CERN to LNGS assuming that the speed of neutrino follows a Galilean transformation where $u=c+V(t)_{E-W-CMB}$ from equation (1). }
\end{figure}
\begin{figure}[h] 
\includegraphics[width=1\textwidth]{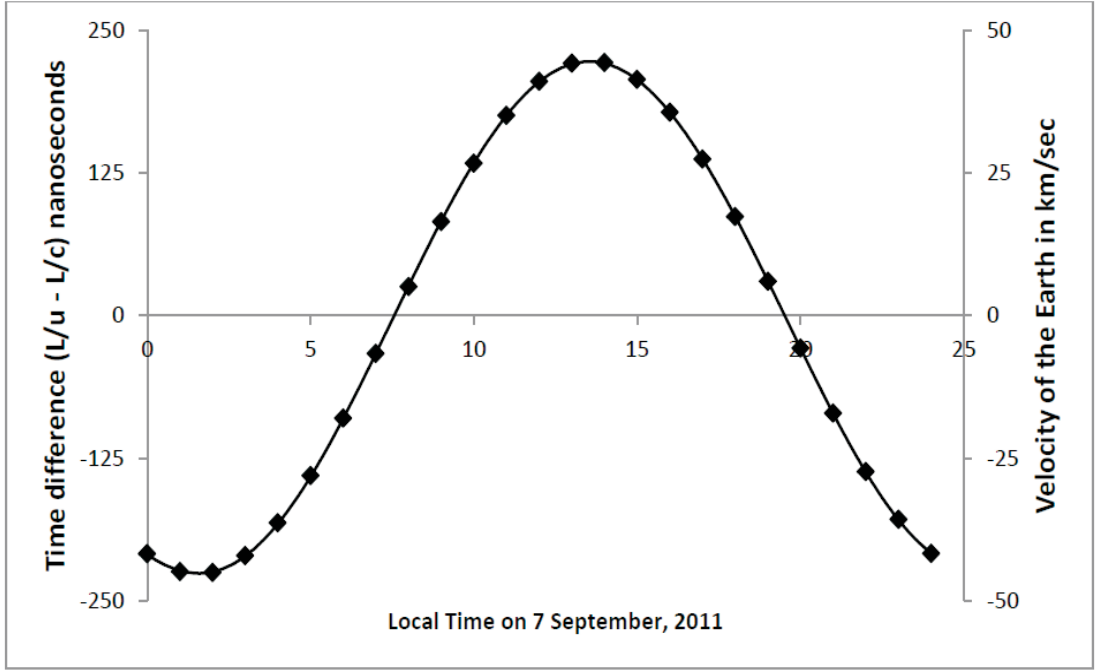} 
\caption{Presentation of a predicted time difference between the time-of-flights corresponding to c and u from CERN to LNGS assuming that the speed of neutrino follows a Galilean transformation where $u=c+V(t)_{E-W}$ from equation (2).}
\end{figure}
\section{Discussion}
The speed of the neutrino is equal to the speed of light in vacuum $(c)$ if the source of neutrinos is attached to a rest frame (say with the CMB). When the source of neutrinos is attached to a moving frame (the Earth), one can assume the speed of the neutrino $u=c+V(t)_{E-W-CMB}$ or $u=c+V(t)_{E-W}$ follows a Galilean transformation. Fig. 3, based on the outcomes of different neutrino experiments, presents an interesting variation of the speed of neutrino which indicates the possibility of our claim. Therefore, one cannot rule out the possibility that the variation is due to the movement of the Earth.  \\
\\ We already presented the time dependent components of the velocities of a laboratory relative to the CMB in equations (1) and in Fig. 2. As we know from our discussion in section 2 that the velocity of the Earth relative to a rest frame is an unresolved problem. Therefore, we derive equation (2), the time dependent velocity of the Earth with respect to the centre of the solar system which was used to analyse classic experiments (such as Ref. [3]) before the invention of the CMB. We presented a predicted time difference between the time-of-flights corresponding to c and u from CERN to LNGS (a baseline distance of 732 km) assuming that the speed of neutrino follows the emission theories. Based on this presentation in Fig. 4 and Fig. 5, we can estimate that the time-difference of 60 nanoseconds reported by the OPERA is within the range of our prediction.    
\section{Conclusions}
We know that the neutrino can escape the core of the star without any incident. It can pass through the Earth without any barrier. As far as we know that almost all of the important properties of neutrinos are still unresolved. All of these peculiar properties of neutrinos will put the outcome of the neutrino experiment in a huge challenge to make it acceptable unambiguously. \\
\\ However, we would like to conclude that while the measured time-difference is consistence with a frame dependent speed of neutrino it is much smaller than the maximum shown in Fig. 4 but the time-difference is comparable to that shown in Fig. 5. This is true of all of the experiments shown in Fig. 3. Thus we think that the time-difference is highly unlikely to be due to a frame dependent speed of the neutrino. Following Maccione [22] we would like to note that there are no sidereal variations that have been measured for OPERA and also the exact time and date of the neutron emission are unclear according to the OPERA report by Adam et al [11]. Also, this is true of all of the experiments shown in Fig. 3. Therefore, we would like to propose that future tests should be made to determine whether there is any sidereal variation in neutrino velocity. Also, two measurements, with six months gap as shown in Fig. 2, can give higher sensitivity to understand the reality.
\section*{Acknowledgements}
The author is indebted to Professor A. D. Stauffer, Department of Physics and Astronomy, York University, Toronto, Canada for valuable comments and discussions. Suggestions of Professor Brendan M. Quine, Department of Earth and Space Science and Space Engineering, York University, Toronto, Canada to consider this 
problem are gratefully acknowledged. This work was supported by York University, Toronto, Canada. Also, the author is thankful to Mr. Nick Balaskas, Department of Physics and Astronomy, York University, Toronto, Canada for his encouragement of this work.     
\section*{References} 
1.     Arya, A. P. (1974) Elementary Modern Physics, (Addison-Wesley,	Physics).\\ 
\\
2.     Jenkins, F. A. and White, H. E. (1987) Fundamentals of Optics (4th edn., M-\indent cGraw Hill International Editions, Physics Series).\\
\\
3.  	Michelson, A.A.  and Morley, E.W. (1887) Am. J. Sci. 34, 333-345. \\
\\
4.	Lorentz, H.A. (1904) Proc. Acad. Sci. Amsterdam, 6, 809. \\
\\
5.	Poincare, H. (1905) C. Rendues Acad. Sci. Paris 140, 1504. \\
\\
6.	Einstein, A. (1905) Ann. Phys. 322, 891-921. \\
\\
7.	Kosteleck,  V.A. and Samuel, S. (1989) Phys. Rev. D 39, 683-685. \\ \\
8.	Amelino-Camelia, G., Ellis, J., Mavromatos, N.E., Nanopoulos, D.V. and Sarkar, S.  \indent (1998) Nature 393, 763-765. \\
\\
9.	Gambini, R.  and Pullin, J. (1999) Phys. Rev. D 59, 124021. \\
\\
10.	Adamson, P. et al. [MINOS Collaboration] (2007) Phys. Rev. D76, 072005. \\ \\
11.	Adam, T., et al. (173 additional authors) (2011) arXiv:1109.4897v2. \\ 
\\
12.	Amelino-Camelia, G., Gubitosi, G., Loret, N., Mercati, F., Rosati, G., and L-\indent ipari, 	P. (2011) arXiv:1109.5172v2. \\
\\
13. Smoot, G.F., Gorenstein, M.V. and Muller, R.A. (1977) Phys. Rev. Lett. 39(14), \indent 898-901. \\
\\
14.	Ahmed, M. F., Quine, B. M., Sargoytchev, S. and Stauffer, A. D.  (2011a) a- \indent ccepted for publication in the Indian Journal of Physics, arXiv:1011.1318v2. \\ \\
15.	Ahmed, M. F., Quine, B. M., Sargoytchev, S. and Stauffer, A. D. (2011b) arXiv:1103.6086v3. \\ \\
16. 	Ahmed, M.F. (2012) PhD thesis, York University, Toronto, Canada (in progress). \\ \\
17.	Elburg, R. A.J. van (2011) arXiv:1110.2685v1. \\ \\
18.	Panofsky, W. K. H. and Phillips, M. (1962) Classical Electricity and Magnetism,	 \indent Second Edition, Addison-Wesley Publishing Company, INC. \\ \\
19.	Zuber, K. (2004) Neutrino Physics, IOP Publishing. \\ \\
20. 	Kalbfleisch, G. R., Baggett, N., Fowler, E. C. and Alspector, J. (1979) Phys. \indent Rev. 	Lett. 43, 1361.\\ \\
21.	Hirata, K. et al. (1987) [KAMIOKANDE-II Collaboration], Phys. Rev. Lett. \indent 58, 	1490. \\ \\
22. Maccione, L., Liberati, S., and Mattingly, D. M. (2011) arXiv:1110.0783v1.
\end{document}